\documentclass{elsart}
\usepackage{epsfig}

\usepackage{amssymb}
\begin{document}

\begin{frontmatter}
\title{Compatibility of QCD sum-rules and Hadron field theory in a dense medium}

\author{R. Aguirre}

\address{Departamento de F\'{\i}sica, Fac. de Ciencias Exactas,\\
Universidad Nacional de La Plata.\\
C. C. 67 (1900) La Plata, Argentina.}

\begin{abstract}
The compatibility of the QCD sum rules and effective hadronic
models predictions are examined. For this purpose we have
considered the results for the nucleon self-energy in a dense
hadronic environment provided by two independent QCD sum-rules
calculations. They are immersed in a theory of hadronic fields
giving rise to non-linear interactions, whose vertices are
parameterized in different ways. Although all of them reproduce
the self-energy used as input, very different descriptions of
nuclear observables are obtained. Only under very definite
circumstances we have found an acceptable agreement with the
nuclear matter properties. To achieve this, phenomenological
parameters are not required at all.
\end{abstract}

\end{frontmatter}

Currently, there is a noticeable interest in interpreting and
formalizing the low energy manifestations of the fundamental
theory of the strong interactions.  However, the mathematical
complexity that Quantum Chromodynamics (QCD) exhibits in this
regime makes impossible to reach this aim. Several approaches were
proposed to circumvent this difficulty, such as lattice
simulations, QCD sum-rules, and effective models, for instance
 Nambu- Jona Lasinio, bag-like models, and chiral perturbation
theory. Each of these treatments emphasizes some aspects of QCD,
considered as the most relevant in the hadronic phase. Contact
with low energy physics is ensured by fixing some model
parameters, such as hadronic masses and decay constants to its
experimental values. It is generally believed that for high
density and/or temperatures, the symmetries of the fundamental
theory would be recovered, therefore effective models
 should reproduce this feature to some degree.\\
The
 low to medium density states of nuclear matter have been
extensively studied by using nuclear potentials. Most of the
nuclear structures, reactions, and dispersion phenomenology have
been successfully described in this way. However, these
formulations lack of Lorentz covariance and they violate causality
when extrapolated to higher densities. The requirements of
covariance and causality can be fulfilled properly within the
hadronic field theory (HFT), whose prototype the Quantum
Hadrodynamics \cite{WALECKA} usually takes dense nuclear matter as
its reference state. This practice contrasts with certain nuclear
potential treatments that refer to in- vacuum scattering lengths,
 ascribing this feature to a more fundamental instance of
nuclear potentials as compared with HFT. Inspired in this
viewpoint, several investigations have been made in order to make
compatible both schemes, due to the simplicity and accuracy of the
HFT treatment. The in-medium self-energy of the nucleon plays a
key role in these attempts, as it is evaluated in both approaches
and then equated to extract medium dependent vertices
\cite{TOKI,HADDAD,LENSKE,TYPEL}.\\
 The
  idea behind this procedure is that when a piece
of information provided by another field of research is coherently
inserted into the framework of the HFT, a simpler and more
intuitive description is obtained instead of stating first
principles interactions and performing involved calculations. This
procedure is specially suited for high density systems since the
details of the interactions are faded out in favor of a
statistical average. A significative exemplification is given by
the  Brown-Rho scaling of hadronic masses. Taking into account the
chiral and scale symmetries of QCD solely, an approximate scaling
law for the in-medium hadronic masses was derived in
\cite{BROWNRHO}. This hypothesis was applied to describe heavy ion
collision, reaching an excellent agreement with the experimental
results for the low mass dilepton production rate \cite{BROWN}.\\
On the contrary, going from low to higher energies, references
\cite{TOKI,HADDAD,LENSKE,TYPEL} take nuclear potentials as the
starting point and results are extrapolated to the medium-high
dense nuclear matter and atomic nuclei. In ref. \cite{LENSKE} it
was stressed that a coherent model requires that the medium
dependence of the above mentioned vertices must be introduced in
terms of hadronic fields. Otherwise, the \emph{rearrangement}
self-energies
are missed and thermodynamical consistency is not satisfied.\\
In this work, we try to match QCD inspired results with the HFT
formalism, by introducing field dependent strength couplings as
proposed in \cite{LENSKE}. For this purpose, we have chosen as
inputs the nucleon self-energies evaluated within the QCD
sum-rules approach \cite{SHIFMAN,IOFFE,DRUKAREV0,COHEN,DRUKAREV}.
We have compared different field parameterizations of the
vertices, and examined the ability of our results to fit the
nuclear matter phenomenology.\\
QCD sum-rules is a powerful procedure to reveal the QCD
foundations of certain hadronic properties. This method relies on
the evaluation of correlation functions in terms of quarks and
gluons degrees, then applying the operator product expansion it is
possible to express them as combinations of perturbative
contributions and condensates (non-perturbative). In the last
step, results are connected with the hadronic counterpart, from
which information on the spectral function can be inferred. The
method was developed to study meson \cite{SHIFMAN} as well as
baryon
 \cite{IOFFE} properties in vacuum, it was subsequently generalized to
study finite density systems \cite{DRUKAREV0,COHEN,DRUKAREV}. We
have used the in-medium nucleon self-energies provided by
references \cite{COHEN,DRUKAREV}, more precisely the simplified
expressions given there as functions of the baryonic density. Both
investigations include non-perturbative contributions representing
one meson exchange, the structure of the nucleon, and multiple
meson exchange. Nevertheless, they differ in the criteria used to
simplify the final formulae. The results of ref. \cite{DRUKAREV}
keep all the contributions averaged over the Borel mass parameter;
whereas, in \cite{COHEN} only the leading term in the operator
product expansion is retained, and the Borel mass is fixed in
order that the effective nucleon mass goes to its experimental
value in vacuum. In the latter case, a linear dependence on the
density is obtained for both the scalar and vector nucleon
self-energy.\\
As it was above mentioned, the Density Dependent Hadron Field
Theory (DDHFT) \cite{LENSKE,TYPEL} attempts to insert the nucleon
self-energy supplied by Dirac-Brueckner calculations with one
boson exchange potentials, into a hadronic field description of
the nuclear matter phenomenology.
 In general terms, we have
adopted the procedure described in ref. \cite{LENSKE} to merge
dynamically the external input into a field model, and we will
indicate where we depart from it.

We start with a lagrangian for nucleons ($\psi$) and scalar and
vector mesons ($\sigma$ and $\omega$ respectively), resembling the
QHD-I model of Walecka \cite{WALECKA} for isospin symmetric
nuclear matter,
\begin{equation}
{\mathcal L}= \bar{\psi}(i \not \! \partial - M + G_s \,\sigma-
G_v \not \! \omega)\, \psi +\frac{1}{2}(\partial_\mu \sigma
\partial^\mu \sigma\,-m_s^2 \sigma^2)-\frac{1}{4}F_{\mu \nu}
F^{\mu \nu}+\frac{1}{2} m_v^2 \omega_\mu \omega^\mu,
\end{equation}
where $F_{\mu \nu}=\partial_\mu \omega_\nu -
\partial_\nu \omega_\mu$ has been used, and the meson masses have been fixed at the
 empirical values
$m_v=783$ MeV, $m_s=550$ MeV. The main difference is that $G_s$
and $G_v$ are not coupling constants as in the Walecka model, but
they are field functionals. Since these vertices are Lorentz
scalars, we expect that the dependence on the fermion fields
appears through the bilinear combinations
$s_1(x)=\bar\psi(x)\psi(x)$, $s_2(x)=\sqrt{n_\mu(x) n^\mu(x)}$,
with $n_\mu(x)=\bar\psi(x)\gamma_\mu \psi(x)$ the baryonic
current. Ref. \cite{LENSKE} only developed the dependence on $s_1$
or $s_2$ ; in this paper, we also consider the eventual dependence
on the meson fields but neglect derivatives of any of the fields,
namely $G_{s,v}=G_{s,v}(s_1,s_2,s_3,s_4)$, with
$s_3=\sqrt{\omega_\mu \omega^\mu}$, and $s_4=\sigma$. The field
equations are obtained in the usual way, by minimizing the action
\begin{eqnarray}
( i\not\!\partial-M+G_s \sigma&-&G_v \not \! \omega+\bar{\psi}
\sum_1^2 \frac{\partial G_s}{\partial s_k}\frac{\partial
s_k}{\partial \bar{\psi}}\sigma -\bar{\psi}\sum_1^2 \frac{\partial
G_v}{\partial s_k}\frac{\partial s_k}{\partial \bar{\psi}}\not \!
\omega
) \psi=0 \label{NUCLEONEQ}\\
\left( \square+m_s^2 \right)\sigma&=&\bar{\psi}\left(
G_s+\frac{\partial G_s}{\partial s_4} \sigma-\frac{\partial
G_v}{\partial
s_4}\not\! \omega \right)\psi \label{SCALAREQ} \\
\partial_\mu F^{\mu \nu}+m_v^2 \omega^\nu&=&\bar{\psi}\left[ G_v \gamma^\nu+\frac{d s_3}{d \omega_\nu}
 \left(-\frac{\partial G_s}
{\partial s_3}\sigma+\frac{\partial G_v}{\partial s_3} \not\!
\omega \right) \right]\psi. \label{VECTOREQ}
\end{eqnarray}

The two last terms within brackets in
Eq. (\ref{NUCLEONEQ}) evaluated in the mean field approximation
can be regarded as the scalar and vector components of the nucleon
self-energies. In order to make the problem solvable, only a set
of variables $s_k$ can be taken into account in the evaluation of
the functionals $G_s$ and $G_v$. We have considered the following
cases: a) both vertices depend only on $s_2$, b) $G_s$ depends on
$s_1$, $G_v$ depends on $s_2$, and c) $G_s$ depends on $\sigma$,
$G_v$ depends on $s_3$. The case (a) has been developed in great
detail in ref. \cite{LENSKE}, the same as case (b) they produce
meson equations resembling those of the Walecka model, and nucleon
self-energies with the \emph{rearrangement} contributions added.
In the instance (c) the self-energies have the same structure as
in QHD-I, but the source term of the meson equations are modified.

The mean field approximation (MFA) is suited to describe static
homogeneous nuclear matter. Within this scheme, meson fields are
replaced by their uniform mean values and bilinear combinations of
spinors are replaced by their expectation values. Therefore within
MFA, Eqs. (\ref{NUCLEONEQ}-\ref{VECTOREQ}) reduce to

\begin{eqnarray}
0&=&(i\not\!\partial-M+\Sigma_s-\not \! \! \Sigma) \psi \label{NUCLEONEQ2}\\
m_s^2 \tilde{\sigma}&=&<\bar{\psi}\,G_s\psi>
+<\bar{\psi}\frac{\partial G_s}{\partial \sigma} \psi \,\sigma>
-<\bar{\psi}\frac{\partial G_v}{\partial
\sigma} \psi\not\! \omega> \label{SCALAREQ2} \\
m_v^2 \tilde{\omega}^\nu&=&<\bar{\psi}\,G_v
\gamma^\nu\psi>-<\bar{\psi}\frac{\partial G_s} {\partial
s_3}\frac{\omega^\nu}{s_3}\psi \,\sigma>
+<\bar{\psi}\frac{\partial G_v}{\partial
s_3}\frac{\omega^\nu}{s_3}\not\! \omega \psi> \label{VECTOREQ2}
\end{eqnarray}

where tildes over the meson symbols stand for their mean field
values, and
\begin{eqnarray}
\Sigma_s&=&<G_s \sigma> + <\bar{\psi}\psi\frac{\partial G_s}{
\partial s_1}\sigma-\bar{\psi}\gamma_\lambda\psi\frac{\partial
G_v}{\partial s_1} \omega^\lambda>,\label{SELFSMFA}\\
\Sigma_\mu&=&<G_v
\omega_\mu>+\frac{1}{s_2}<(-\bar{\psi}\psi\frac{\partial G_s}{
\partial s_2}\sigma+\bar{\psi}\gamma_\lambda\psi\frac{\partial
G_v}{\partial s_2} \omega^\lambda) \bar{\psi}\gamma_\mu \psi>
.\label{SELFVMFA}
\end{eqnarray}
These results are simplified within the assumptions (a-c) above,
for the sake of completeness we consider separately each of these
cases:
\begin{eqnarray}
\mbox{a)} \,\Sigma_s&=&G_s \tilde{\sigma},\, \Sigma_\mu=
G_v\tilde{\omega}_\mu+(dG_v/dn)\,\tilde{\omega}^\nu
\Lambda_{\nu\mu}-(dG_s/dn)\,\tilde{\sigma}\Lambda_\mu,\,
\tilde{\sigma}=G_s n_s/m_s^2, \nonumber\\
\tilde{\omega}_\mu&=&G_v n_\mu/m_v^2,\mbox{with}
\;\Lambda_{\nu\mu}=<\bar{\psi} \gamma_\nu \psi
\bar{\psi}\gamma_\mu \psi>/n,\, \mbox{and}\;
\Lambda_\mu=<\bar{\psi} \psi
\bar{\psi}\gamma_\mu \psi>/n, \nonumber \\
\mbox{b)}\, \Sigma_s&=&\tilde{\sigma}(G_s+ n_s \; dG_s/dn_s ),\,
\Sigma_\mu= G_v\tilde{\omega}_\mu+(dG_v/dn)\,\tilde{\omega}^\nu
\Lambda_{\nu\mu},\nonumber \\
&& \,\tilde{\sigma}, \,\mbox{and}
\;\tilde{\omega}_\mu\, \mbox{have the same expressions as in (a)}
,\nonumber\\
\mbox{c)}\, \Sigma_s&=&G_s \tilde{\sigma}, \,\Sigma_\mu=G_v
\tilde{\omega}_\mu,\; m_s^2 \tilde{\sigma}=G_s
n_s+\tilde{\sigma}n_s (dG_s/d\sigma), \nonumber\\&&m_v^2
\tilde{\omega}_\mu=G_v n_\mu+ (dG_v/d\tilde{\omega})\, n_\nu
\tilde{\omega}^\nu \tilde{\omega}_\mu/s_3,\nonumber
\end{eqnarray}
 we have
used $G_{s,v}=<G_{s,v}>$, $n_s=<\bar{\psi}\psi>$,
 $n_\mu=<\bar{\psi}\gamma_\mu\psi>$, $n=n_0$ is the baryonic
density, and we have taken into account the fact that $s_2=n$, as
it is clear when it is evaluated in the reference system of rest
matter. Finally, we have introduced the simplifying assumption
that $G_{s,v}$ and its derivatives can be extracted from the
expectation values, even when they are field functionals.
Regarding  to the expectation values $\Lambda_{\mu\nu}$ and
$\Lambda_\mu$, they can be evaluated by using the Wick theorem and
the propagator for the nucleons in the medium, giving rise to
direct and exchange terms. In our calculations we have neglected
the Fock terms, in which case we have obtained
$\Lambda_{\mu\nu}=n_\nu n_\mu/n$, $\Lambda_\mu=n_s n_\mu/n$.

In previous investigations \cite{TOKI,HADDAD,LENSKE}, the
procedure to define the functionals $G_s$ and $G_v$ consisted of
including the nucleon self-energies as input into the equations
\begin{equation}\Sigma_s^{inp}=G_s n_s/m_s^2, \;\;\; \Sigma_\mu^{inp}=G_v
n_\mu/m_v^2, \label{DDHFT}
\end{equation}

where $\Sigma_s^{inp},\Sigma_\mu^{inp}$ are the self-energies
obtained with one boson exchange potentials in the
Dirac-Brueckner-Hartree-Fock approach for symmetric nuclear
matter. The Eqs. (\ref{DDHFT}) coincide with Eqs.
(\ref{SELFSMFA})-(\ref{SELFVMFA}) if the \emph{rearrangement}
terms were omitted, in spite of the importance of these
contributions for satisfying the thermodynamic consistency, as
stressed in ref. \cite{LENSKE}. At this point we deviate from the
procedure of the DDHFT in two aspects; firstly, we adopt as input
the nucleon self-energies $\Sigma^{\mbox{\tiny QCDSR}}$ grounded
on the fundamental theory of strong interactions. Secondly, we
insert this input into the full expressions (\ref{SELFSMFA}) and
(\ref{SELFVMFA}), in fact we regard from the very beginning the
field dependence of $G_{s,v}$. Therefore, the vertices are the
solution of a differential equation which must be solved together
with the self-consistent condition for $\tilde{\sigma}$ in cases
a) and b), or a pair of differential equations for
$\tilde{\sigma}$ and $\tilde{\omega}$ in the case c). The explicit
form of these solutions are shown below, for each of the instances
considered
\begin{eqnarray}
\mbox{a)}\,G_s&=&\frac{\Sigma_s}{\tilde{\sigma}}^{\mbox{\tiny
QCDSR}}, \, G_v^2=2\,\left(\frac{ m_v}{n}\right)^2\int_0^n dn'
\left[\Sigma_v^{\mbox{\tiny
QCDSR}}+\left(\frac{n_s}{m_s}\right)^2G_s\frac{dG_s}{dn}\right],\nonumber
\\
\mbox{b)}\,G_s^2&=&2\, \left(\frac{m_s}{n_s}\right)^2\int_0^n \,
dn'\frac{dn_s}{dn}\Sigma_s^{\mbox{\tiny QCDSR}},\,G_v^2=2\,
\left(\frac{m_v}{n}\right)^2\int_0^n \, dn'\,\Sigma_v^{\mbox{\tiny
QCDSR}},\nonumber \\
\mbox{c)}\,\tilde{\sigma}^2&=&\frac{2}{m_s^2}\int_0^n \, dn'\,n_s
\frac{d\Sigma_s}{dn}^{\mbox{\tiny
QCDSR}},\,\tilde{\omega}^2=\frac{2}{m_v^2}\int_0^n \, dn'\,n'
\frac{d\Sigma_v}{dn}^{\mbox{\tiny QCDSR}},\nonumber
\end{eqnarray}
where we used that the spatial components of tetravectors become
null for isotropic matter, and we considered
$\Sigma_{s,v}^{\mbox{\tiny QCDSR}}$ as being parameterized in
terms of only the baryonic density $n$. The expressions for
$G_{s,v}$ become singular at zero density, nevertheless they have
finite limits: $G_s^2\rightarrow m_s^2 d\Sigma_s^{\mbox{\tiny
QCDSR}}(0)/dn, \,G_v^2\rightarrow m_v^2 d\Sigma_v^{\mbox{\tiny
QCDSR}}(0)/dn$, for all the cases under consideration.

\begin{figure}[t]\vspace{-3cm}
\begin{center}
\psfig{file=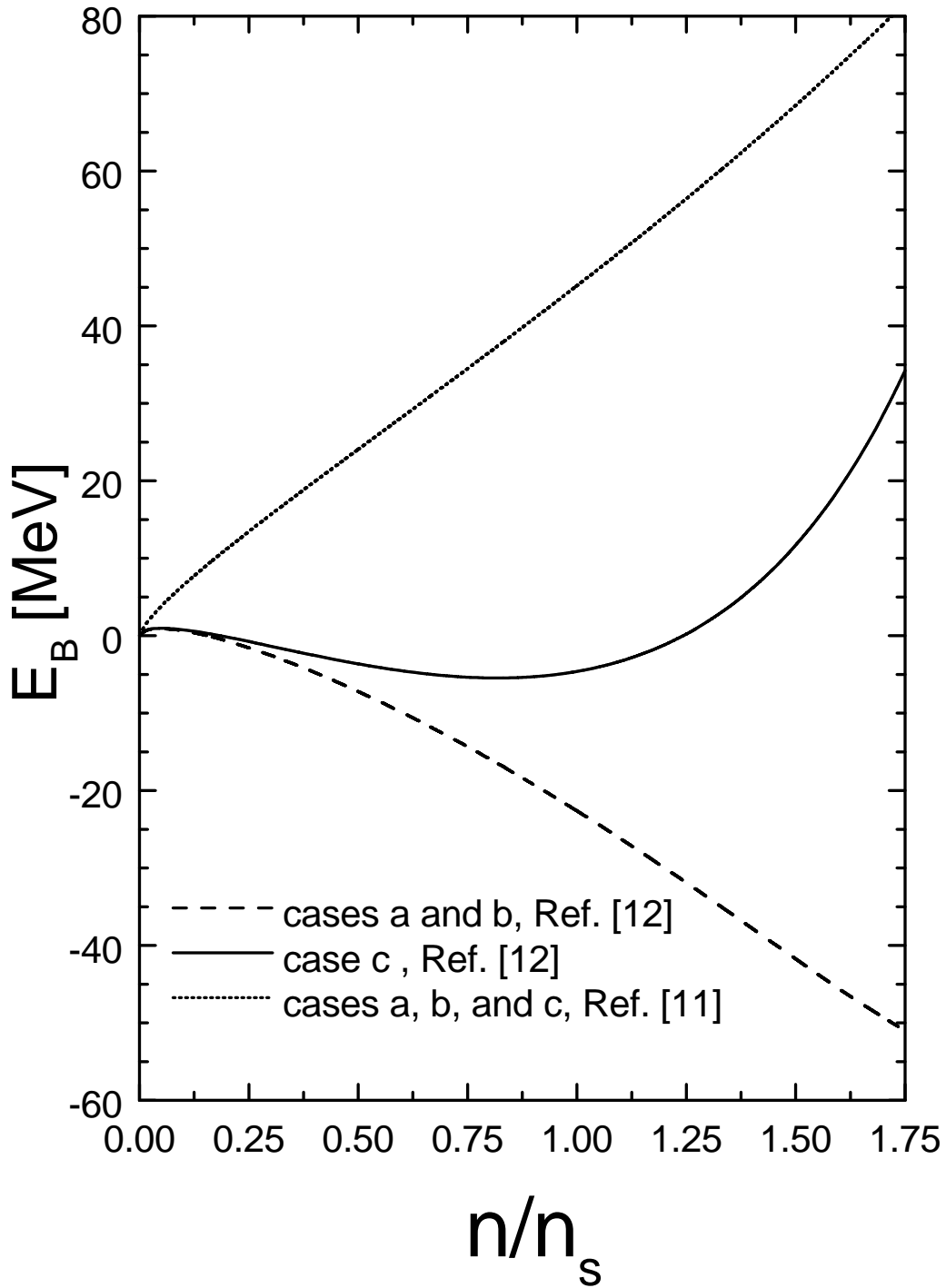,width=\textwidth}\end{center} \caption{The
binding energy as a function of the baryonic density for each of
the parameterizations of the interaction vertices, cases a-c. By
using as input the self-energy given by ref. \cite{COHEN} all the
curves coincide, in the case of including the nucleon self-energy
of ref. \cite{DRUKAREV}  the case c) distinguishes clearly from
the other ones. The line convention is indicated in the
figure.}\label{FIG1}
\end{figure}

\begin{figure}[t]\vspace{-3cm}
\begin{center}
\psfig{file=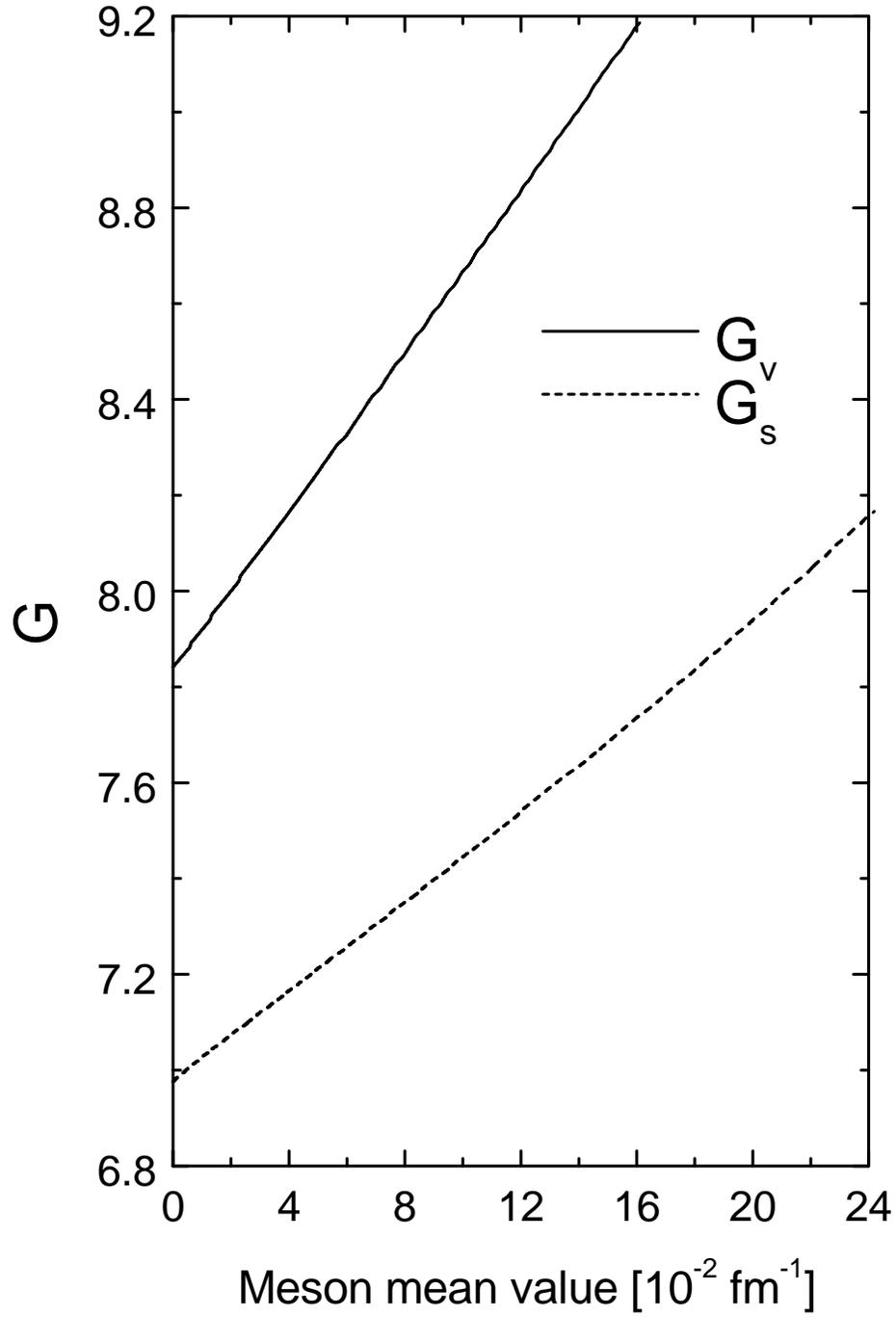,width=\textwidth}\end{center} \caption{The
interaction vertices of the hadronic field model as functions of
the meson mean field values. The model reproduces the nucleon
self-energy given by ref. \cite{DRUKAREV}. } \label{FIG2}
\end{figure}

Up to this point, we have constructed a hadronic effective
interaction capable of reproduce the nucleon self-energies as they
were evaluated in the QCD sum-rules; therefore, our findings are
subjected to the same limitations in applicability.\\ In ref.
\cite{WEISE}, a similar purpose was attained by adopting the
results of \cite{COHEN} and  simulating the hadronic interaction
by means of four fermion couplings. The vertices are taken as
functions of only $s_2$, they were determined in the same way as
in the DDHFT, namely, using Eq. (\ref{DDHFT}). Nevertheless, the
authors consider that the effect of chiral fluctuations is
predominant in the many body aspects of nuclear matter, and the
variation of the self-energies according to QCD sum-rules
calculations has a
stabilization effect at high densities.\\
In the next step we have evaluated the energy-momentum tensor
$T^{\mu \nu}$ by the canonical procedure, and finally, the energy
per unit volume in the MFA is obtained by taking the in-medium
expectation value of $T^{00}$. This gives

\begin{eqnarray}
E_{MFA}&=&\int_0^{p_F} \frac{d^3p}{(2 \pi)^3}\sqrt{p^2+M^{*
2}}+n_s(\Sigma_s^{\mbox{\tiny
QCDSR}}-G_s\tilde{\sigma})+
\frac{1}{2}(m_s^2\tilde{\sigma}^2+m_v^2\tilde{\omega}^2),
\end{eqnarray}

where we have introduced the Fermi momentum $p_F$, related to the
baryonic density by $n=2p_F^3/(3 \pi^2)$, and the effective
nucleon mass $M^*=M-\Sigma_s^{\mbox{\tiny QCDSR}}$. It must be
noticed that the sign of the scalar self-energy has been changed
as regards the one used in refs. \cite{COHEN,DRUKAREV}, in
order to adopt the practices of QHD models. \\
Using
 the thermodynamical
relation $P=\mu n-E_{MFA}$ we have obtained the pressure of the
nuclear matter, with the chemical potential given by
$\mu=E_F+\Sigma_v$ for cases a) and b), and $\mu=E_F+n \,
d\Sigma_v/dn$ for the remaining one, being $E_F=\sqrt{p_F^2+M^{*
2}}$.\\
 A
feature of isospin symmetric nuclear matter is that the energy per
particle has a minimum, giving rise to bound states at zero
temperature. This is the least requirement that a model of nuclear
matter should satisfy. A measure of this property is the binding
energy $E_B=E_{MFA}/n-M$, which is well determined to have a
minimum value $E_B\simeq -16$ MeV at the normal density $n_s=0.15
fm^{-3}$. We have examined the binding energy in the three cases
studied for the functional dependence of $G_{s,v}$, which are
displayed
 in Fig. \ref{FIG1} in terms of the baryonic density.
 As we can see, none of these instances is able to fit the
 saturation properties of nuclear matter. Our results with the
 input taken from ref. \cite{COHEN} does not distinguish
 among the different parameterizations of the vertices; on the
 other hand, the case c) evaluated with the self-energies given by ref. \cite{DRUKAREV}
 (case cD in the following) clearly stands out from the
 other ones. The non-monotonic behavior of the curve cD is more compatible with the
 phenomenology, more precisely the
 binding energy has a minimum $E_B=-4$ MeV at the density
 $n=0.8 \, n_s$, not far from the empirical saturation
 point.\\
 It is apparent that the approaches a) and b)
 yield undistinguishable results for $E_B$.  On the
 other hand, the case c) produces higher meson mean values,
 enhancing the energy density. This fact reverses the decreasing
 trend of $E_B$ obtained for the cases a) and b) in the right hand side of
 this figure, the net  behavior is non-monotonous as shown there.
 On the other hand, the same effect applied to the increasing
 curves (a-b) on the left panel, reinforces the growth. Although,
 the small increment is not appreciable within the scale of this
 figure.\\
 It has been stated that in order to obtain the slight binding energy and simultaneously
 a high nuclear spin-orbit potential, the scalar and vector components of the
 nucleon self-energy must have nearly the same value, of roughly three hundred
 MeV for densities  about $n_s$ \cite{SEROT}. This requisite is
 satisfied by both refs. \cite{COHEN} and \cite{DRUKAREV}; however, as
 can be deduced from the results a) and b) it is not sufficient
 to  determine conclusively the many body aspects of the system.\\
 As the QCD
 sum-rules calculations are subjected to uncertainties of diverse origins, we
 have introduced two independent scaling factors $X_s$ and $X_v$ for
 each of the nucleon self-energies used as inputs, and we have explored
 the range $\mid X_{s,v}-1 \mid<0.25$. We have assumed these factors
 independent of the density, consequently their derivatives are scaled by the
 same magnitude. We have found that within this range of
 variation, only the case (cD)  is able to reproduce simultaneously the
 binding energy and the density of saturation with a slight scaling
 $X_s\simeq 1.19,\,X_v\simeq 1.12$. Aiming to a qualitative description for all the
 densities considered, we have introduced a uniform scaling of the
 self-energies $\Sigma_{s,v}^{\mbox{\tiny QCDSR}}(n) \rightarrow X_{s,v}
 \Sigma_{s,v}^{\mbox{\tiny QCDSR}}(n)$ within the scheme c).\\
 As the next step, we have examined the behavior of the Landau
 parameters $F_0$ and $F_1$ for symmetric nuclear matter. They can be evaluated
 as the Legendre projections of the second derivative of $E_{MFA}$
 with respect to the baryonic density, for a deduction see
 \cite{MATSUI}. These parameters are very useful in regard to
 collective phenomena of the dense nuclear environment, as for
 instance, the giant monopolar and quadrupolar modes, or the sound
 velocity \cite{KURASAWA}. The nuclear compressibility can be
 evaluated as $K=3p_F^2(1+F_0)/E_F$, its value at the
 equilibrium density is constrained experimentally to range between
230 MeV $<K<$ 270 MeV. Within the approach (cD) we have obtained
$K=245$ MeV, and related to it is the giant monopole mode energy
$E_M=\omega_0/A^{1/3}$. In our calculations we have obtained
$\omega_0=108$ MeV, whereas, the experience yields
$\omega_0^{exp}=80$ MeV. On the other hand the scaled calculations
lead to  the excessively high values $K=660$ MeV, and
$\omega_0=180$ MeV. Another interesting property is the excitation
energy of the quadrupole state, given by the relation
$E_Q=\omega_2/A^{1/3}$ with
$\omega_2=p_F\sqrt{2}/(1.2\mu\sqrt{1+F_1/3})$, yielding in our
calculations $\omega_2=78$ MeV, and $\omega_2=100$ MeV in the
non-scaled and scaled cases, respectively. These values must be
compared with the empirical result $\omega_2^{exp}=63$ MeV. On the
other hand, the numerical values obtained for $\omega_0$ and
$\omega_2$ without scaling, can be well accommodated among the
theoretical results for the isoscalar giant resonances in the
non-linear models of HFT \cite{CENTELLES}.
 Finally, the sound velocity $v_s=p_F^2(1+F_0)/(3\mu
E_F)$ takes on the value $v_s/c=0.17$ and $v_s/c=0.28$ at the
saturation density for the non-scaled and scaled instances,
respectively. The magnitude of $v_s$ increases with density
quickly, becoming acausal before $n \approx 2 n_s$ is reached.
This feature puts severe limits to the range of
applicability of the scheme.\\
In Fig. \ref{FIG2} we show the vertices $G_{s,v}$ for the (cD)
approach, in terms of their respective variables. They are well
fitted by the expressions $G_s=a/(1+b\, s_3)$ and $G_v=A+B\, s_4$,
with $a=7$, $b=-0.59 fm^{-1}$, $A=7.83$, and $B=8.39 fm^{-1}$.

To sum up, in the present work we have examined the possibility of
matching QCD sum-rules predictions and hadronic field theories. A
successful combination would result in a profitable feedback, by
giving foundations to hadronic models and corroborating
simultaneously the QCD sum-rules methods and findings. In
particular we have used results for the nucleon self-energy in a
dense hadronic environment derived in two independent treatments
of QCD sum-rules, with diverse degree of approximation. As a first
step, we have studied different formal schemes of merging these
inputs into hadronic field models and we have checked their
ability to fit the basic nuclear matter phenomenology. We have
introduced interaction vertices $G_s$ and $G_v$, which are
functionals of the hadronic fields, and we have determined them by
requiring the reproduction of the self-energies used as input. It
must be emphasized that in this procedure we have included the
full expression for the self-energies, having obtained a set of
differential equations for the definition of these vertices,
instead of algebraic ones,.
 We have found that only a parameterization of the vertices in terms
of mesonic degrees is compatible with the basic features of dense
nuclear matter. In addition we have presented a fitting in terms
of rational functions, which make evident the non-linear quality
of the meson-nucleon coupling. If we allow an uncertainty margin
of $20\%$ in the sum-rules calculations, this scheme is able to
fit precisely the saturation properties of the isospin symmetric
nuclear matter. Another physical quantities, which are related to
collective nuclear phenomena, agree with the order of magnitude of
the experimental values, although they are not admissible as a
quantitative description.\\
It must be stressed that no adjustable parameters are introduced,
only the QCD sum-rules self-energies are required as input. The
parameters $X_s, X_v$ were introduced only to study the
versatility of the results.

 We conclude that compatibility of QCD
sum-rules and hadronic field models is feasible. However, a
detailed qualitative description will require further improvements
in QCD sum-rules calculations, in particular the high density QCD
phenomenology should be properly included. Therefore, additional
refinements and higher order calculations will be useful and
deserve to be investigated. Improved results could be inserted
into hadronic models to test another significative manifestations
of the hadronic phase such as  finite nuclei and the eventual
deconfinement phase transition.

\newpage
\newpage

\end{document}